\documentclass[11pt]{article}
\usepackage{amsmath}
\usepackage{amsfonts}
\usepackage{amssymb}
\usepackage{theorem}

\def\p{\partial}
\newtheorem{prop}{Proposition}

\newcommand{\be}{\begin{equation}}
\newcommand{\ee}{\end{equation}}
\newcommand{\bea}{\begin{eqnarray}}
\newcommand{\eea}{\end{eqnarray}}
\newcommand{\beaa}{\begin{eqnarray*}}
\newcommand{\eeaa}{\end{eqnarray*}}

\newcommand{\nn}{\nonumber}
\renewcommand{\d}{\mathrm{d}}
\begin{document}
\title{On a class of reductions of Manakov-Santini hierarchy
connected with the interpolating system}
\author{
L.V. Bogdanov\thanks
{L.D. Landau ITP, Kosygin str. 2,
Moscow 119334, Russia, e-mail
leonid@landau.ac.ru}}
\maketitle
\begin{abstract}
Using Lax-Sato formulation of Manakov-Santini hierarchy, 
we introduce a class of reductions, 
such that zero order  reduction of this class corresponds to
dKP hierarchy, and the first order reduction gives the hierarchy associated
with the interpolating system
introduced by Dunajski. We present Lax-Sato form of reduced hierarchy for 
the interpolating system
and also for the reduction of arbitrary order. Similar to dKP hierarchy, Lax-Sato
equations for $L$ (Lax fuction) due to the reduction 
split from Lax-Sato equations for $M$ (Orlov function), and the reduced hierarchy
for arbitrary order of reduction is defined by Lax-Sato equations for $L$ only.
Characterization of the class of reductions in terms of the dressing data is given.
We also consider a waterbag reduction of the interpolating system hierarchy, 
which defines (1+1)-dimensional systems of hydrodynamic type. 
\end{abstract}
\section{Introduction}
In this work we construct a class of reductions of the hierarchy associated
with the system recently introduced by Manakov and Santini \cite{MS06}
(see also \cite{MS07}, \cite{MS08}),
\bea
u_{xt} &=& u_{yy}+(uu_x)_x+v_xu_{xy}-u_{xx}v_y,
\nn\\
v_{xt} &=& v_{yy}+uv_{xx}+v_xv_{xy}-v_{xx}v_y,
\label{MSeq}
\eea
whose Lax pair is
\bea
&&
\partial_y\mathbf{\Psi}=((p-v_{x})\partial_x - u_{x}\partial_p)\mathbf{\Psi},
\nn\\
&&
\partial_t\mathbf{\Psi}=((p^2-v_{x}p+u -v_{y})\partial_x
-(u_{x}p+u_{y})\partial_p)\mathbf{\Psi},
\label{MSLax}
\eea
where $p$ plays a role of a spectral variable. Manakov-Santini system is a 
generalization of dispersionless KP
(Khohlov-Zabolotskaya) equation to the case of general (non-Hamiltonian)
vector fields in the Lax pair. For $v=0$  
the system reduces to the dKP equation.
Respectively, $u=0$ reduction gives an equation 
\cite{Pavlov03} (see also \cite{Dun04,MS02,MS04})
\begin{equation}
 v_{xt} = v_{yy}+v_xv_{xy} - v_{xx}v_y.
\label{Pavlov}
\end{equation}

Using Lax-Sato formulation of the hierarchy \cite{BDM06,BDM07,LVB09}, 
we introduce a class of reductions, 
such that zero order  reduction of this class corresponds to
dKP hierarchy, and the first order reduction gives the hierarchy connected
with the interpolating system, which was introduced in \cite{Dun08},
where it was proved that it is "the most general symmetry reduction of the
second heavenly equation by a 
conformal Killing vector with a null self-dual derivative".
In \cite{Dun08} it was also shown that the interpolating system corresponds to
simple differential reduction $cu=bv_x$ of Manakov-Santini equation.
We present Lax-Sato form of reduced hierarchy for interpolating system
and also for reduction of arbitrary order. Similar to dKP hierarchy, Lax-Sato
equations for $L$ (Lax fuction) due to the reduction 
split from Lax-Sato equations for $M$ (Orlov function), and the reduced hierarchy
for arbitrary order of reduction is defined by Lax-Sato equations for $L$ only.
In terms of Manakov-Santini system this class defines 
differential reductions (not changing
the number of dimensions). Characterization
of the class of reductions in terms of the dressing data is given. 
We also consider waterbag type reductions of reduced
hierarchies (including interpolating equation hierarchy), which define
(1+1)-dimensional systems of hydrodynamic type.

Reductions of Manakov-Santini system were considered also in the works
\cite{BChang08}, \cite{Chang09}, \cite{PavlovChang09}, 
concentrating mostly on (1+1)-dimensional reductions
of hydrodynamic type.

\section{Manakov-Santini hierarchy}
Manakov-Santini hierarchy is defined by Lax-Sato equations \cite{BDM06,BDM07,LVB09}
\bea
&&
\frac{\partial}{\partial t_n}\begin{pmatrix}
L\\
M
\end{pmatrix}=
\left(\left(
\frac{ L^n L_p}{\{L,M\}}\right)_+
{\partial_x}
-\left(\frac{ L^n L_x}{\{L,M\}}\right)_+
{\partial_p}\right)\begin{pmatrix}
L\\
M
\end{pmatrix},
\label{genSato1}
\eea 
where $L$, $M$, corresponding to Lax and Orlov functions of dispersionless KP hierarchy,
are the series
\bea
&&
L=p+\sum_{n=1}^\infty u_n(\mathbf{t})p^{-n},
\label{form01}
\\&&
M=M_0+M_1,\quad M_0=\sum_{n=0}^\infty t_n L^{n},
\nn \\&&
M_1=\sum_{n=1}^\infty v_n(\mathbf{t})L^{-n}=
\sum_{n=1}^\infty \tilde v_n(\mathbf{t})p^{-n},
\label{form1}
\eea
and $x=t_0$, $(\sum_{-\infty}^{\infty}u_n p^n)_+
=\sum_{n=0}^{\infty}u_n p^n$, $\{L,M\}=L_pM_x-L_xM_p$.
A more standard choice of times for dKP hierarchy corresponds to 
$M_0$=${\sum_{n=0}^\infty (n+1)t_n L^{n}}$, it is easy to transfer to it 
by rescaling of times.

Lax-Sato equations  (\ref{genSato1}) are equivalent to the generating relation
\cite{BDM06,BDM07,LVB09}
\be
\left(\frac{\d L\wedge \d M}{\{L,M\}}\right)_-=0,
\label{analyticity0}
\ee
where differential takes into account all times $\mathbf{t}$ and 
variable $p$.

Equations (\ref{genSato1}) imply that
the dynamics of the Poisson bracket $J=\{L,M\}$ is described by the equation
\cite{BChang08}
\bea
&&
\frac{\partial}{\partial t_n}\ln J=
\left(
A_n
{\partial_x}
-
B_n{\partial_p}
\right)
\ln J +\p_x A_n
-\p_p B_n,
\label{genSato2}
\\&&
A_n=\left(\frac{ L^n L_p}{J}\right)_+,\quad 
B_n=\left(\frac{ L^n L_x}{J}\right)_+.
\nn
\eea 
This equation together with the first equation of (\ref{genSato1}) forms
a closed system which defines Manakov-Santini hierarchy and 
can be used as an equivalent to system (\ref{genSato1}), it is very useful
for the description of reductions. Thus, to define Manakov-Santini hierarchy,
it is possible to consider the equations
\bea
\frac{\partial}{\partial t_n}L&=&
\left(\left(
L^n L_pJ^{-1}\right)_+
{\partial_x}
-\left({ L^n L_x}{J}^{-1}\right)_+
{\partial_p}\right)L,
\nn\\
\frac{\partial}{\partial t_n}\ln J&=&
\left(\left(
{ L^n L_p}{J}^{-1}\right)_+
{\partial_x}
-\left({ L^n L_x}{J}^{-1}\right)_+
{\partial_p}\right)\ln J 
\nn\\&&
+ \p_x\left(
{ L^n L_p}{J}^{-1}\right)_+ -\p_p \left({ L^n L_x}{J}^{-1}\right)_+\quad
\label{genSato22}
\eea 
for the series $L(p)$ (\ref{form01}) and $J$,
\bea
J=1+\sum_1^{\infty}j_n(\mathbf{t}) L^{-n}= 1+\sum_1^{\infty}\tilde j_n(\mathbf{t}) p^{-n}.
\label{formJ}
\eea
Function $M$ can be found from $L$ and $J$
using the relation \cite{BChang08}
\beaa
J={\{L,M\}}=(\p_p L)\p_x M|_L,
\eeaa
where $|_L$ means that a partial derivative is taken for fixed $L$. Then
\bea
\p_x M|_L=J(\p_p L)^{-1}=J \p_L p(L),
\label{MJ}
\eea
and, introducing series for $p(L)$ (inverse to $L(p)$ (\ref{form01})),
\bea 
p=L + \sum_1^{\infty}p_n(\mathbf{t}) L^{-n},
\label{pL}
\eea
it is possible to find coefficients of the series for $\p_x M|_L$ explicitly and define
the function $M$. For the first coefficient of the series (\ref{form1}) we get
$\p_x v_1(\mathbf{t})=j_1(\mathbf{t})$. In the case of Hamiltonian vector fields $J=1$
and $\p_x M|_L=\p_L p(L)$.

Lax-Sato equations for the first two flows of the hierarchy (\ref{genSato1})
\bea
&&
\partial_y
\begin{pmatrix}
L\\
M
\end{pmatrix}
=((p-v_{x})\partial_x - u_{x}\partial_p)
\begin{pmatrix}
L\\
M
\end{pmatrix}
\label{MSLax1},
\\
&&
\partial_t
\begin{pmatrix}
L\\
M
\end{pmatrix}
=((p^2-v_{x}p+u -v_{y})\partial_x
-(u_{x}p+u_{y})\partial_p)
\begin{pmatrix}
L\\
M
\end{pmatrix}
\label{MSLax2},
\eea
where $u=u_1$, $v=v_1$, $x=t_0$, $y=t_1$, $t=t_2$,
correspond to the Lax pair (\ref{MSLax}) of Manakov-Santini system (\ref{MSeq}).

Equation (\ref{MSLax1}) gives recursion relations, defining the coefficients
of the series $L(p)$, $M(p)$ (\ref{form01}),
(\ref{form1}) through the functions $u$, $v$,
\bea
&&
\partial_x u_{n+1}=\partial_y u_n + v_x u_n - (n-1)u_x u_{n-1},
\label{recurL}
\\
&&
\partial_x \tilde v_{n+1}-u_n=\partial_y \tilde v_n + v_x \tilde v_n - 
(n-1)u_x \tilde v_{n-1},\quad n\geqslant 1,\;\tilde v_1=v.
\eea
Using these relations, Manakov-Santini system can be directly obtained 
from equation (\ref{MSLax2}) without the application of compatibility
conditions for linear equations. It is also possible to use equations
for $\ln J$ (\ref{genSato22}), the first two flows read
\bea
&&
\partial_y
\ln J
=((p-v_{x})\partial_x - u_{x}\partial_p)
\ln J -v_{xx}
\label{MSLax1J},
\\
&&
\partial_t
\ln J
=((p^2-v_{x}p+u -v_{y})\partial_x
-(u_{x}p+u_{y})\partial_p)
\ln J -v_{xx}p-v_{xy},\qquad 
\label{MSLax2J}
\eea 
and recursion relation for $\ln J= \sum_{n=1}^\infty (\ln J)_n p^{-n}$ 
is similar to recursion for $L(p)$,
\beaa
&&
\partial_x (\ln J)_{n+1}=\partial_y (\ln J)_n + v_x (\ln J)_n - (n-1)u_x (\ln J)_{n-1},
\eeaa
where $n\geqslant 1$, $(\ln J)_1=v_x$.

\section{A class of reductions connected with the interpolating system}
In this section we consider a class of reductions of Manakov-Santini hierarchy,
characterized by existense of order $k$ polynomial (with respect to $p$) solution
of non-homogeneous linear equation (\ref{genSato2}). For $k=0$ this reduction
corresponds to Hamiltonian vector fields and dKP hierarchy. 
For $k=1$ we obtain the interpolating system \cite{Dun08} hierarchy. 
For general $k$ $J$ can be explicitely expressed through $L$, and the reduced 
hierarchy is defined by Lax-Sato equations for $L$ only (similar to dKP hierarchy).

Let $\ln J$ satisfy non-homogeneous equations (\ref{genSato2}) and $L$ satisfy
homogeneous equations (\ref{genSato1}), than the function $\ln J+F(L)$ also satisfies
equations (\ref{genSato2}). We define a class of reductions of Manakov-Santini hierarchy
by the condition
\be
(\ln J - \alpha L^k)_-=0,
\label{red00}
\ee
where $\alpha$ is a constant,
that means that equations (\ref{genSato2}) have an analytic solution 
$(\ln J - \alpha L^k)$.
This condition defines a reduction because $A_n$, $B_n$ in equations (\ref{genSato2})
are polynomials, and the dynamics, defined by these equations, preserves analitycity
of the functions, so analytic solutions form an invariant manifold. Thus, if
$(\ln J - \alpha L^k)(x,p)$ is polynomial with respect to $p$ 
at initial point in higher times, then it is
polynomial for arbitrary values of higher times. 

Reduction (\ref{red00}) is completely characterized by the existence of
polynomial solution of equations (\ref{genSato2}).
\begin{prop}
Existence of polynomial solution 
$$
f=-\alpha p^k +\sum_0^{i=k-2} f_i(\mathbf{t})p^i,
$$
(where coefficients $f_i$ don't contain constants, see below)
of equations (\ref{genSato2}),
\bea
&&
\frac{\partial}{\partial t_n}f=
\left(
A_n
{\partial_x}
-
B_n{\partial_p}
\right)
f +\p_x A_n
-\p_p B_n,
\label{gS}
\eea
is equivalent to the reduction condition (\ref{red00}).
\end{prop}
\textbf{Proof} First, reduction condition (\ref{red00}) directly implies that
$f=(\ln J - \alpha L^k)$ is a polinomial solution of equations (\ref{gS})
of required form, that proves that existence of polynomial solution is necessary.

To prove that it is sufficient, we note that $F=\ln J -f$ solves homogeneous equations
(\ref{gS}) (equations (\ref{genSato1})).
Let us expand $p$ into the powers of $L$ (\ref{pL}),
and represent $F$ in the form
$$
F=\alpha L^k + \sum_{-\infty}^{i=k-2} F_i(\mathbf{t})L^i,
$$
where $F_i(\mathbf{t})$ can be expressed through $f_i(\mathbf{t})$ and 
coefficients of expansion of $J$ and $L$ (respectively,
$j_n (\mathbf{t})$ and $u_n (\mathbf{t})$). It is easy to check that $F$ solves 
homogeneous equations (\ref{gS}) 
iff all the coefficients $F_i(\mathbf{t})$ are constants.
Suggesting that coefficients $f_i$ of the polynomial $f(p)$ don't contain constants
(in the sense that they are equal to zero if all the coefficients $j_n=u_n=0$),
we come to the conclusion that $\ln J -f=\alpha L^k$. \hfill$\square$\\

The simplest case $k=0$ corresponds to Hamiltonian vector fields. Indeed, in this case
$J=1$, and from equations (\ref{genSato2}) we have 
$$
\p_x A_n-\p_p B_n=0.
$$

In the case $k=1$
\bea
&&
(\ln J - \alpha L)_-=0\Rightarrow (\ln J - \alpha L)=(\ln J - \alpha L)_+=-\alpha p,
\nn\\&&
J=\exp\alpha(L-p).
\label{red1}
\eea
So, similar to the case of Hamiltonian vector fields, equation for $L$ splits
and the reduced hierarchy is defined by Lax-Sato equations
\bea
\frac{\partial}{\partial t_n}L=(e^{\alpha(p-L)}L^n L_p)_+\p_x L-
(e^{\alpha(p-L)}L^n L_x)_+\p_p L.
\label{red1h}
\eea
Generating relation for the reduced hierarchy reads
\beaa
\left(e^{\alpha(p-L)}{\d L\wedge \d M}\right)_-=0,
\eeaa
or, equivalently,
\beaa
\left(e^{-\alpha L}{\d L\wedge \d M}\right)_-=0.
\eeaa
Representing  relation (\ref{red1}) as a series in $p^{-1}$, in the first nontrivial
order we get (see (\ref{MJ}))
\be
\alpha u=j_1=v_x,
\label{reduv}
\ee
that is exactly the condition used in \cite{Dun08} to reduce Manakov-Santini system 
to interpolating equation
($\alpha=\frac{c}{b}$ in the notations of \cite{Dun08}). 
Manakov-Santini system (\ref{MSeq}) with reduction (\ref{reduv}) is equivalent
to interpolating equation up to a simple transformation, and we will call hierarchy
(\ref{red1h}) {\em the interpolating equation hierarchy}.

Reduction condition (\ref{red1}) implies that $(-\alpha p)$ is a solution of equations
(\ref{genSato2})
(in fact, these conditions are {\em equivalent}), 
and, substituting it, we get reduction equations in term of
vector fields components,
\bea
\p_x A_n
-\p_p B_n-B_n=0.
\label{vectorred}
\eea
It is easy to check that for $n=1$ we obtain a reduction condition
(\ref{reduv}). 
\subsection*{General $k$}
In the general case,
\bea
&&
(\ln J - \alpha L^k)_-=0\Rightarrow (\ln J - \alpha L^k)
=(\ln J - \alpha L^k)_+=-\alpha (L^k)_+,
\nn
\\&&
J=\exp\alpha(L^k - (L^k{}_+))=\exp\alpha(L^k{}_-),
\label{redk1}
\eea
and Lax-Sato equations of reduced hierarchy read
\bea
\frac{\partial}{\partial t_n}L=(e^{-\alpha(L^k{}_-)}L^n L_p)_+\p_x L-
(e^{-\alpha (L^k{}_-)}L^n L_x)_+\p_p L.
\label{red1hk}
\eea
These equations  imply equations (\ref{genSato22}) for $J$ (\ref{redk1}),
function $M$ is defined by relation (\ref{MJ}),
\beaa
\p_x M|_L=J(\p_p L)^{-1}=e^{\alpha(L^k - (L^k{}_+))}(\p_p L)^{-1}.
\eeaa
Generating relation (\ref{analyticity0}) in this case takes the form
\bea
\left(e^{-\alpha L^k}{\d L\wedge \d M}\right)_-=0.
\label{analyticity0k}
\eea
Reduction (\ref{redk1}) is equivalent to the condition that $(-\alpha L^k{}_+)$
is a solution to equations (\ref{genSato2}), that gives a differential
characterization of reduction in terms of Manakov-Santini hierarchy,
\bea
&&
\frac{\partial}{\partial t_n}(\alpha L^k{}_+)=
\left(
A_n
{\partial_x}
-
B_n{\partial_p}
\right)
(\alpha L^k{}_+) -\p_x A_n
+\p_p B_n,
\label{genSato2red}
\\&&
A_n=\left(\frac{ L^n L_p}{J}\right)_+,\quad 
B_n=\left(\frac{ L^n L_x}{J}\right)_+.
\nn
\eea 
For the first flow $n=1$ we obtain a condition (compare (\ref{MSLax1J}))
\bea
&&
\partial_y
(\alpha L^k{}_+)
=((p-v_{x})\partial_x - u_{x}\partial_p)
(\alpha L^k{}_+) +v_{xx}.
\label{MSLax1Jred}
\eea
This condition defines a differential reduction of Manakov-Santini system.

Let us consider in more detail the case $k=2$. Reduction is defined by relation 
(\ref{redk1}),
\be
J=e^{\alpha(L^2{}_-)}
\label{redk12}.
\ee
Taking an expansion into powers of $p^{-1}$,
in the first nontrivial order we get
\beaa
j_1=2\alpha u_2.
\eeaa
Using recursion formula (\ref{recurL}), we obtain
\beaa
\p_x u_2=u_y+v_x u_x.
\eeaa
Thus we come to the conclusion that in terms
of Manakov-Santini system (\ref{MSeq}) reduction (\ref{redk12}) leads to a condition
\be
2\alpha(u_y+v_x u_x)=v_{xx}
\label{MSredk22}
\ee
This condition defines a differential reduction of Manakov-Santini system.

Another way to obtain differential reduction is to use relation (\ref{MSLax1Jred}).
Indeed, $(L^2{}_+)=p^2+2u$, and, substituting this expression to relation
(\ref{MSLax1Jred}),
we get
\beaa
&&
2\alpha u_y
=2\alpha((p-v_{x})u_x - u_{x}p)+v_{xx}\Rightarrow 2\alpha (u_y+2v_x u_x)=v_{xx}.
\eeaa
Relation (\ref{MSLax1Jred}) explicitly gives differential reductions of arbitrary
order $k$ for Manakov-Santini system.

For illustration we will also calculate differential reduction of Manakov-Santini
system of the order $k=3$. In this case $(L^3{}_+)=p^3+3pu+3u_2$, and,
substituting this expression to (\ref{MSLax1Jred}), we get
\be
3\alpha\left(
\p_y(u_y+u_x v_x)+\p_x(u_y v_x+u_x v_x^2+u u_x)\right)=v_{xxx}.
\label{k3red}
\ee
\subsection*{A pair of reductions with different $k$ --
reduction to (1+1)}
If we consider a pair of reductions with different $k$, we obtain a closed
(1+1)-dimensional system of equations for the functions $u$, $v$.
First let us consider reductions of interpolating system, i.e.,
reduction with $k=1$, which leads to the condition (\ref{reduv}),
together with reduction (\ref{red00}) of some order $k\neq 1$ (with a constant $\beta$).

For $k=2$, using  (\ref{red00}) and (\ref{MSredk22}), we obtain a 
system
\beaa
&&
u_y+v_x u_x=(2\beta)^{-1}v_{xx},
\\&&
v_x=\alpha u,
\eeaa
which implies hydrodynamic type equation
(Hopf type equation) for $u$,
$$
u_y+ \alpha u u_x=\frac{\alpha}{2\beta}u_{x}.
$$

The system for $k=3$ reads (see (\ref{k3red}))
\beaa
&&
\p_y(u_y+u_x v_x)+\p_x(u_y v_x+u_x v_x^2+u u_x)={3\beta}^{-1}v_{xxx},
\\&&
v_x=\alpha u,
\eeaa
it implies an equation for $u$,
$$
u_{yy}+
\p_x(2\alpha u_y u+\alpha^2 u_x u^2+u u_x-\frac{\alpha}{3\beta}u_{x})=0,
$$
which can be rewritten as a system of hydrodynamic type for two functions
$u$, $w$,
\beaa
&&
w_y =
(\frac{\alpha}{3\beta} -\alpha^2 u^2 -u)u_{x}-2\alpha u w_x,
\\&&
u_y=w_x.
\eeaa
A system of equations of hydrodynamic type corresponding to the reduction of 
interpolating system of arbitrary order $k>3$ can be obtained
using the observation that
$f=\beta L^k{}_+-\alpha p$ is a solution of linear equation
$$
\p_y f=(p-\alpha u)\p_x f - u_x\p_p f,
$$
which provides a system of hydrodynamic type for the coefficients
of the polynomial $f=\beta p^k + k \beta u p^{k-2} -\alpha p+ 
\sum_{i=0}^{k-3} f_i p^i$, namely
\beaa
\p_y u&=& (k\beta)^{-1}\p_x f_{k-3}- \alpha u \p_x u,
\\
\p_y f_{k-3} &=& \p_x f_{k-4}- \alpha u \p_x f_{k-3} -k(k-2)\p_x u ,
\\
\p_y f_{i} &=&\p_x f_{i-1}  - \alpha u \p_x f_{i} -(i+1)f_{i+1}\p_x u,
\quad 0<i<k-3,
\\
\p_y f_{0}&=&- \alpha u\p_x f_{0}-(f_{1}-\alpha)\p_x u.
\eeaa

Let us also consider a simple example of a system
defined by two reductions of higher order, taking reductions
of the order 2 (\ref{MSredk22}) and of the order 3 (\ref{k3red}),
\beaa
&&
u_y+v_x u_x=(2\alpha)^{-1}v_{xx},
\\&&
\left(
\p_y(u_y+u_x v_x)+\p_x(u_y v_x+u_x v_x^2+u u_x)\right)=(3\beta)^{-1}v_{xxx}.
\eeaa
This system can be rewritten as a system of hydrodynamic type for the functions
$u$, $w=v_x$,
\beaa
&&
u_y+w u_x=(2\alpha)^{-1}w_x,
\\&&
w_y=\frac{2\alpha}{3\beta}w_x - ww_x-2\alpha u u_x.
\eeaa
\section{A waterbag reduction for the interpolating system hierarchy}
For the class of reduced hierarchies defined by Lax-Sato equations
(\ref{red1hk}) it is possible to consider manifold of solutions of the form
\bea
L(p,x)=p-\sum_{i=1}^N c_i \ln(p-w_i(x)),\quad \sum_{i=1}^N c_i=0,
\label{waterbag}
\eea
where $c_i$ are some constants.
Due to the fact that coefficients of vector fields  in equations (\ref{red1hk})
are polynomial, and 'plus' projection of equations is identically zero by construction,
it is straightforward to demonstrate that this manifold is invariant under dynamics,
so it defines a reduction (this type of reduction is known for dKP hierarchy
as a waterbag reduction). Each of 
Lax-Sato equations (\ref{red1hk}) in this case is equivalent to the closed 
(1+1)-dimensional system of equations for the functions $u_i$.

Let us study in more detail the waterbag reduction for 
interpolating equation 
hierarchy (\ref{red1h}).
First two Lax-Sato equations of the hierarchy read
\bea
&&
\p_y L=(p-\alpha u)\p_x L - u_x\p_p L,
\nn\\&&
\p_t L=(p^2-\alpha u p- \alpha u_2 +u)\p_x L -
(u_x p - \alpha u u_x + \p_x u_2)\p_p L.
\label{Laxwaterbag}
\eea
For Lax-Sato function of the form (\ref{waterbag}) the coefficients  of
expansion $u_n$ are expressed through the functions $w_i$ as
\bea
u_n=\sum_{i=1}^N \frac{c_i}{n} w_i^n,
\label{coeffwaterbag}
\eea
Substituting ansatz (\ref{waterbag}) to Lax-Sato equations (\ref{Laxwaterbag})
and using formula (\ref{coeffwaterbag}), we obtain two closed (1+1)-dimensional
systems of equations for the functions $w_i$,
\bea
\p_y w_i&=&\left(w_i -\alpha \sum_{i=1}^N {c_i} w_i\right)\p_x w_i + 
\p_x\sum_{i=1}^N {c_i} w_i,
\nn\\
\p_t w_i&=&\left(w_i^2-\alpha w_i\sum_{i=1}^N {c_i} w_i- 
\alpha \sum_{i=1}^N \frac{c_i}{2} w_i^2 +
\sum_{i=1}^N {c_i} w_i\right)\p_x w_i 
\nn\\&&\quad
+
\left(w_i-\alpha \sum_{i=1}^N {c_i} w_i\right) \p_x \sum_{i=1}^N {c_i} w_i
+ \p_x \sum_{i=1}^N\frac{c_i}{2} w_i^2.
\label{LaxwaterbagH}
\eea
These systems (as well as higher flows) are compatible, because they are constructed
as a reduction of the flows of Manakov-Santini hierarchy to the invariant manifold
(\ref{waterbag}). On the invariant manifold equations (\ref{LaxwaterbagH}) are equivalent
to Lax-Sato equations of Manakov-Santini hierarchy. Equations (\ref{LaxwaterbagH})
are (1+1)-dimensional systems of hydrodynamic type, their common solution gives
a solution of interpolating equation (Mananakov-Santini system (\ref{MSeq}) with
the reduction $\alpha u=v_x$)
by the formula
\beaa
u=\sum_{i=1}^N {c_i} w_i.
\eeaa
In the case $\alpha=0$ formulae (\ref{LaxwaterbagH}) give the waterbag reduction
of the dKP hierarchy \cite{BKwaterbag} (to match (\ref{LaxwaterbagH}) to the formulae
of the work \cite{BKwaterbag}, it is necessary to rescale the times).

Minimal number of components $w_i$ in equations (\ref{LaxwaterbagH}) is two,
and for the simplest case $N=2$, 
$L(p,x)=p-c\ln\frac{p-w_1(x)}{p-w_2(x)}$, an explicit form of hydrodynamic
type system corresponding to the first flow of (\ref{LaxwaterbagH}) is
\beaa
&&
\p_y w_1=\p_x\left(\frac{1}{2}w_1^2 +c(w_1-w_2)\right)- 
\alpha c(w_1-w_2)\p_x w_1,
\\&&
\p_y w_2=\p_x\left(\frac{1}{2}w_2^2 +c(w_1-w_2)\right)- 
\alpha c(w_1-w_2)\p_x w_2,
\eeaa
and the second flow reads
\beaa
\p_t w_1&=&\p_x\left(\frac{1}{3}w_1^3
+c w_1(w_1-w_2)
+\frac{c}{2} (w_1^2-w_2^2)
\right)
\\&&\quad
-\alpha\left(c w_1(w_1-w_2)\p_x w_1+\frac{c^2}{2}\p_x (w_1-w_2)^2\right),
\\
\p_t w_2&=&\p_x\left(\frac{1}{3}w_2^3
+c w_2(w_1-w_2)
+\frac{c}{2} (w_1^2-w_2^2)
\right)
\\&&\qquad
-\alpha\left(c w_2(w_1-w_2)\p_x w_2+\frac{c^2}{2}\p_x (w_1-w_2)^2\right).
\eeaa
Zakharov reduction, corresponding to rational $L$ with simple poles,
can be considered as a degenerate case of the waterbag reduction, when 
pairs of  functions $w_i$ coincide.  In the two-component case,
considering the limit 
$c\rightarrow \infty$, $w_1-w_2=c^{-1}u$, we get $L=p+\frac{u}{p-w}$,
and the equations of reduced hierarchy can be obtained as a limit of equations for the
waterbag reduction. For the first two flows
\beaa
&&
\p_y w=\p_x\left(\frac{1}{2}w^2 +{u}\right)- 
\alpha {u}\p_x w,
\\&&
\p_y {u}=\p_x\left(w{u}\right)- 
\alpha {u}\p_x {u},
\eeaa
and
\beaa
\p_t w&=&\p_x\left(\frac{1}{3}w^3
+ 2w{u}
\right)
-\alpha\left(w{u}\p_x w+\frac{1}{2}\p_x {u}^2\right),
\\
\p_t {u}&=&\p_x\left(
w^2{u}+{u}^2
\right)
-\alpha{u}\p_x(w {u}).
\eeaa
A common solution of these systems gives
a solution $u$ of interpolating equation. 
\section{Characterization of reductions
in terms of the
dressing data}
A dressing scheme for Manakov-Santini hierarchy can be formulated
in terms of two-component nonlinear Riemann problem on the unit circle $S$
in the complex plane of the variable $p$,
\bea
L_\text{in}=F_1(L_\text{out},M_\text{out}),
\nn\\
M_\text{in}=F_2(L_\text{out},M_\text{out}),
\label{RiemannMS}
\eea
where the functions 
$L_\text{in}(p,\mathbf{t})$, $M_\text{in}(p,\mathbf{t})$ 
are analytic inside the unit circle,
the functions $L_\text{out}(p,\mathbf{t})$, $M_\text{out}(p,\mathbf{t})$ 
are analytic outside the
unit circle and have an expansion of the form (\ref{form01}), (\ref{form1}).
The functions $F_1$, $F_2$ are suggested to define (at least locally) 
diffeomorphism of the plane, $\mathbf{F}\in\text{Diff(2)}$, and we call them
dressing data. It is straightforward to demonstrate that the problem
(\ref{RiemannMS}) implies analyticity of the differential form
$$
\Omega_0=\frac{\d L\wedge \d M}{\{L,M\}}
$$
(where differential takes into account all times $\mathbf{t}$ and $p$)
in the complex plane and generating relation (\ref{analyticity0}), thus defining
a solution of Manakov-Santini hierarchy. Considering a reduction to area-preserving
diffeomorphisms \text{SDiff(2)}, we obtain the dKP hierarchy. 

To obtain interpolating system, it is necessary
to consider a more general class of reductions.
Let $G_1(\lambda,\mu)$, $G_2(\lambda,\mu)$ define an area-preserving
diffeomorphism, $\mathbf{G}\in\text{SDiff(2)}$, 
$$
\left|\frac{D(G_1,G_2)}{D(\lambda,\mu)}\right|=1.
$$
Let us fix a pair of analytic functions $f_1(\lambda,\mu)$, $f_2(\lambda,\mu)$
(reduction data)
and consider a problem
\bea
f_1(L_\text{in},M_\text{in})=
G_1(f_1(L_\text{out},M_\text{out}),f_2(L_\text{out},M_\text{out})),
\nn\\
f_2(L_\text{in},M_\text{in})=G_2
(f_1(L_\text{out},M_\text{out}),f_2(L_\text{out},M_\text{out})),
\label{Riemannred}
\eea
which defines a reduction of MS hierarchy.
In terms of initial Riemann problem for MS hierarchy  (\ref{RiemannMS}),
which can be written in the form
\be
(L_\text{in},M_\text{in})=\mathbf{F}(L_\text{out},M_\text{out}),
\label{RiemannMSbis}
\ee
the reduction condition for the dressing data reads
\be
\mathbf{f}\circ\mathbf{F}\circ\mathbf{f}^{-1}\in \text{SDiff(2)}.
\label{redMSdressing}
\ee
In terms of equations of
MS hierarchy the reduction is characterized by the condition
$$
(\d f_1(L,M)\wedge \d f_2(L,M))_\text{out}=
(\d f_1(L,M)\wedge \d f_2(L,M))_\text{in},
$$
thus the form 
$$
\Omega_\text{red}=\d f_1(L,M)\wedge \d f_2(L,M)
$$
is analytic in the complex plane, and  reduced hierarchy is defined by
the generating relation
$$
(\d f_1(L,M)\wedge \d f_2(L,M))_-=0.
$$
Taking 
\bea
f_1(L,M)&=&L,
\nn\\
f_2(L,M)&=&e^{-\alpha L^n}M,
\label{reddiffMS}
\eea
we obtain the generating relation
\beaa
\left(e^{-\alpha L^k}{\d L\wedge \d M}\right)_-=0,
\eeaa
coinciding with (\ref{analyticity0k}).
Thus we come to the following conclusion:
\begin{prop}
A class of reductions (\ref{red00})  is characterized in terms of 
the dressing data for the
problem (\ref{RiemannMSbis}) 
by the condition (\ref{redMSdressing}), where $\mathbf{f}$ is defined by the
formulae (\ref{reddiffMS}). 
\end{prop}
For interpolating equation $f_1=L$, $f_2=e^{-\alpha L}M$, and the Riemann
problem (\ref{Riemannred}) can be written in the form
\beaa
L_\text{in}&=&
G_1(L_\text{out},e^{-\alpha L_\text{out}}M_\text{out}),
\\
M_\text{in}&=&e^{\alpha G_1(L_\text{out},e^{-\alpha L_\text{out}}M_\text{out})}
G_2
(L_\text{out},e^{-\alpha L_\text{out}}M_\text{out}),
\eeaa
where $\mathbf{G}\in \text{SDiff(2)}$.
\section*{Acknowledgments}
The author is grateful to S.V. Manakov and M.V. Pavlov for
useful discussions.
This research was partially supported by the Russian Foundation for
Basic Research under grants no. 06-01-89507 
(Russian-Taiwanese grant 95WFE0300007),
08-01-90104, 07-01-00446, 09-01-92439, and by the President of Russia
grant 4887.2008.2 (scientific schools).

\end{document}